\documentclass[aps,pra,superscriptaddress,twocolumn,showpacs]{revtex4}

\usepackage{graphics}
\usepackage{epsfig}
\usepackage{graphicx}
\usepackage{amsthm} 
\usepackage{amsbsy} 
\usepackage{dcolumn}   
\usepackage{verbatim}   
\usepackage{color}       
\usepackage{subfigure}  
\usepackage{amsmath,amsfonts,amssymb,graphics,graphicx,epsfig,color,natbib}

\theoremstyle{remark}

\theoremstyle{definition}

\newcommand{\ie}{{\it i.e.\,}}

\newcommand{\tr}{\mathrm{tr}\,} 
\newcommand{\bra}[1]{\left\langle#1\right|} 
\newcommand{\ket}[1]{\left|#1\right\rangle} 
\newcommand{\proj}[1]{\ket{#1}\!\bra{#1}}

\newcommand{\icfo}{\affiliation{ICFO--Institut de Ci\`encies Fot\`oniques, E-08860 Castelldefels, Barcelona, Spain}}
\newcommand{\icrea}{\affiliation{ICREA--Instituci\'o Catalana de Recerca i Estudis Avan\c{c}ats, Lluis Companys 23, 08010 Barcelona,
Spain}}

\newcommand{\uniroma}{\affiliation{Dipartimento di Fisica, Universit\`{a} Sapienza di Roma, Roma 00185, Italy}}
\newcommand{\msdf}{\affiliation{Museo Storico della Fisica e Centro Studi e Ricerche Enrico Fermi, Via Panisperna 89/A, Compendio del Viminale, 00184 Roma, Italy}}
\newcommand{\ino}{\affiliation{Istituto Nazionale di Ottica (INO-CNR), L.go E. Fermi 6, I-50125 Florence, Italy}}
\newcommand{\sevilla}{\affiliation{Departamento de F\'{\i}sica Aplicada II, Universidad de Sevilla, 41012 Sevilla, Spain}}
\newcommand{\estocolmo}{\affiliation{Department of Physics, Stockholm University, S-10691 Stockholm, Sweden}}

\begin{document}

\title{Fully nonlocal quantum correlations}
\author{Leandro Aolita}\icfo
\author{Rodrigo Gallego}\icfo
\author{Antonio Ac\'in}\icfo \icrea
\author{Andrea Chiuri}\uniroma
\author{Giuseppe Vallone}\uniroma \msdf
\author{Paolo Mataloni}\uniroma \ino
\author{Ad\'an Cabello}\sevilla \estocolmo

\date{\today}

\begin{abstract}
Quantum mechanics is a nonlocal theory, but not as nonlocal as the
no-signalling principle allows. However, there exist quantum
correlation that exhibit maximal nonlocality: they are as
nonlocal as any nonsignalling correlations and thus have a local
content, quantified by the fraction $p_L$ of events admitting  a
local description, equal to zero. We exploit the known link between the Kochen-Specker and Bell 
theorems to derive a maximal violation of a Bell inequality from every Kochen-Specker proof. We then show that these Bell inequalities
lead to experimental bounds on the local content of quantum
correlations that are significantly better than those based on
other constructions. We perform the experimental demonstration of
a Bell test originating from the Peres-Mermin Kochen-Specker
proof, providing an upper bound on the local content $p_L\lesssim
0.22$.

\end{abstract}
\maketitle


\section{Introduction}
Since the seminal work by Bell \cite{bell}, we know that there
exist quantum correlations that cannot be thought of
classically. This impossibility is known as nonlocality and
follows from the fact that the correlations obtained when performing
local measurements on entangled quantum states may violate a Bell
inequality, which sets conditions satisfied by all classically correlated systems.

The standard nonlocality scenario consists of two distant systems
on which two observers, Alice and Bob, perform respectively $m_a$ and  $m_b$ different
measurements of $d_a$ and $d_b$  possible outcomes. The outcomes of Alice
and Bob are respectively labeled $a$ and $b$, while their
measurement choices are $x$ and $y$, with $a=1,\ldots,d_a$, $b=1,\ldots,d_b$, $x=1,\ldots,m_a$, and
$y=1,\ldots,m_b$. The correlations between the two systems are
encapsulated in the joint conditional probability distribution
$P(a,b|x,y)$.

This probability distribution should satisfy the no-signalling
principle, which states that no faster-than-light communication
is possible. When the measurements by  the two observers define spacelike separated events, this implies that the
marginal distributions for Alice (Bob) should not depend on
Bob's (Alice's) measurement choice, \ie $\sum_b
P(a,b|x,y)=P(a|x),\,\forall\ y$, and similarly for Bob. These
linear constraints define the set of nonsignalling
correlations. Quantum correlations in turn are those that can be
written as $P(a,b|x,y)=\tr(\rho_{AB} M^x_a\otimes M^y_b)$,
where $\rho_{AB}$ is a bipartite quantum state and $M^x_a$ and
$M^y_b$ define local measurements by the observers. Finally,
classical correlations are defined as those that can be written
as
$P(a,b|x,y)=\sum_{\lambda}p(\lambda)P_A(a|x,\lambda)P_B(b|y,\lambda)$.
These correlations are also called local, as outcome $a$ ($b$)
is locally generated from input $x$ ($y$) and the
pre established classical correlations $\lambda$.

The violation of Bell inequalities by entangled states implies that the set
of quantum correlations is strictly larger than the classical one.
A similar gap appears when considering quantum
versus general nonsignalling correlations: there exist
correlations that, despite being compatible with the no-signalling
principle, cannot be obtained by performing local measurements
on any quantum system \cite{pr}. In particular, there exist
nonsignalling correlations that exhibit stronger nonlocality, in the sense of giving larger Bell violations, than any
quantum correlations [see Fig. 1 (a)].


Interestingly, there are situations in which this second gap
disappears: quantum correlations are then maximally nonlocal,
as they are able to attain the maximal Bell violation
compatible with the no-signalling principle. Geometrically, in
these extremal situations quantum correlations reach the border
of the set of nonsignalling correlations [see
Fig. 1 (b)]. From a quantitative point of view,
it is possible to detect this effect by computing the local
fraction \cite{epr2} of the correlations. This quantity measures
the fraction of events that can be described by a local model. Given $P(a,b|x,y)$,
consider all possible decompositions,
\begin{equation}\label{epr2dec}
P(a,b|x,y)=q_L P_L(a,b|x,y) + (1-q_L) P_{NL}(a,b|x,y),
\end{equation}
in terms of arbitrary local and nonsignalling distributions,
$P_L(a,b|x,y)$ and $P_{NL}(a,b|x,y)$, with respective weights
$q_L$ and $1-q_L$, where $0\leq q_L\leq1$. The local fraction of
$P(a,b|x,y)$ is defined as the maximum local weight over all
possible decompositions as \eqref{epr2dec}:
\begin{equation}
\label{p_LDEF}
 p_L\doteq\max_{\{P_L,P_{NL}\}}q_L.
\end{equation}
It can be understood as a
measure of the nonlocality of the correlations. Maximally
nonlocal correlations feature $p_L=0$ [see Fig. 1 (b)].

\begin{figure}
\begin{center}
  \includegraphics[width=8cm,angle=0]{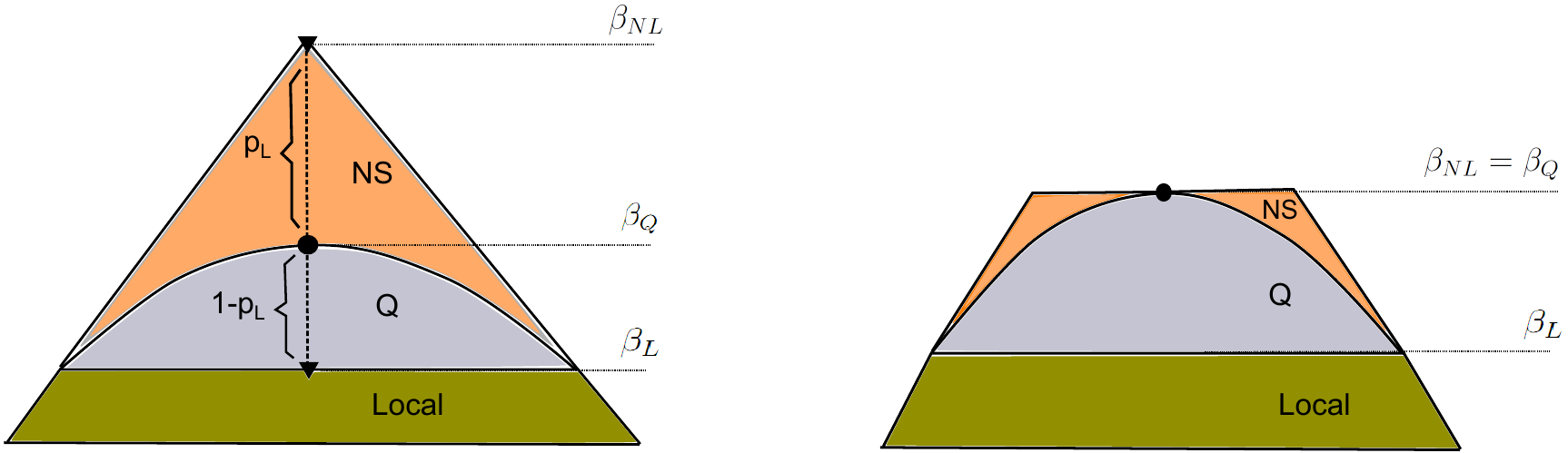}
 \caption{\textbf{Nonsignalling, quantum, and classical
correlations}. The set of nonsignalling correlations defines a polytope. The set of
quantum correlations is contained in the set of
nonsignalling correlations. The set of classical correlations
is also a polytope and is contained inside the quantum set. (a) In general, the set of quantum correlations is not
tangent to the set of nonsignalling correlations. This means
that the maximal value $\beta_Q$ of a Bell inequality
achievable by quantum correlations is above the local bound
$\beta_L$ but strictly smaller than the maximal value
$\beta_{NS}$ for nonsignalling correlations. (b) In the present Bell tests, in contrast, quantum
correlations are tangent to the set of nonsignalling
correlations and thus attain the nonsignalling value of a
Bell inequality. The corresponding upper bound on the local
fraction is zero, which
discloses the full nonlocal nature of quantum mechanics.}
\label{correlations}
\end{center}
\end{figure}

Any Bell violation provides an upper bound on the local
fraction of the correlations that cause it. In fact, a Bell
inequality is defined as $\sum
T_{a,b,x,y}P(a,b|x,y)\leq\beta_L$, where $T_{a,b,x,y}$ is a
tensor of real coefficients. The maximal value of the left-hand
side of this inequality over classical correlations defines the
local bound $\beta_L$, whereas its maximum over quantum and
nonsignalling correlations gives the maximal quantum and
nonsignalling values $\beta_Q$ and $\beta_{NL}$, respectively.
From this and \eqref{epr2dec} it follows immediately that \cite{bkp}
\begin{equation}\label{plbound}
 p_L\leq\frac{\beta_{NL}-\beta_Q}{\beta_{NL}-\beta_L}\doteq{p_L}_{max}.
\end{equation}
Thus, quantum correlations feature $p_L=0$ if (and, in fact, only
if) they violate a Bell inequality as much as any nonsignalling
correlations.


In this work we study the link between the Kochen-Specker
(KS) \cite{ks} and Bell's theorems, previously considered in
Refs.~\cite{refs,avncabello,pseudo,freewill,horo}. We recast this
link in the form of Bell inequalities maximally violated by
quantum states. We then show that the resulting Bell inequalities
can be used to get experimental bounds on the nonlocal content of
quantum correlations that are significantly better than Bell
tests based on more standard Bell inequalities or multipartite
Greenberger-Horne-Zeilinger (GHZ) paradoxes~\cite{ghz}. This
allows us to perform an experimental demonstration, which yields
an experimental upper bound on the local part
${p_L}_{max}=0.218\pm0.014$. To our knowledge, this represents the
lowest value ever reported, even taking into account multipartite 
Bell tests.
\section{GENERAL FORMALISM}

In this section, we present the details of the
construction to derive different Bell inequalities
maximally violated by quantum mechanics from every proof of the KS theorem. This construction was first
introduced in~\cite{refs} and was later applied in the
context of ``all-versus-nothing'' nonlocality
tests~\cite{avncabello}, pseudo-telepathy games (see~\cite{pseudo}
and references therein), the free-will theorem~\cite{freewill},
and quantum key distribution~\cite{horo}. Here we exploit it to
generate quantum correlations with no local part.

Recall that the KS theorem studies whether deterministic outcomes
can be assigned to von Neumann quantum measurements, in contrast
to the quantum formalism which can only assign probabilities. A
von Neumann measurement $z$ is defined by a set of $d$ orthogonal
projectors acting on a Hilbert space of dimension $d$. Consider
$m$ such measurements, given by $m\times d$ rank-1 projectors
$\Pi^z_i$, with $z=1,\ldots,m$ and $i=1,\ldots,d$, such that
$\Pi^z_i\Pi^z_{i'}=\delta_{i,i'}$ and $\sum_i\Pi^z_i=\openone$ for
all $z$, with  $\openone$ being the identity operator. The theorem
studies maps from these measurements to deterministic $d$-outcome
probability distributions. In addition, an
extra-requirement is imposed on the maps: the assignment has to be noncontextual. That
is, if a particular outcome, corresponding to a projector
$\Pi^z_i$, is assigned to a given measurement, the same outcome
must be assigned to all the measurements in which this projector
appears. Formally, this means that the assignment map, denoted by
$v_A$, acts actually on projectors rather than on measurements:
$v_A(\Pi^z_i)\in\{0,1\}$,  such that $\sum_i v_A(\Pi^z_i)=1$ for
all $z$. The KS theorem shows that noncontextual deterministic
assignments do not exist.

Although this impossibility follows as a corollary of Gleason's
theorem \cite{Gleason57}, one virtue of the proofs of the KS
theorem \cite{ks,Mermin90, Peres91,CEG96} is that they involve a
finite number of measurements. More precisely, each KS proof
consists of a set of $m$ measurements (contexts) as above but
chosen so as to share altogether $p$ projectors $\tilde\Pi_j$,
with $j=1,\ldots,p$, that make noncontextual deterministic
assignments incompatible with the measurements' structure. Denote
by $D_{j}$ the set of two-tuples $D_{j}=\{(i,z)\}$ such that
$(i,z)\in D_{j}$ if $\Pi_{i}^{z}=\tilde{\Pi}_{j}$. Each set of
two-tuples $D_{j}$ collects the indexes of all common projectors
among all different measurements.

\begin{figure}
 \begin{center}
  \includegraphics[width=8.3cm,angle=0]{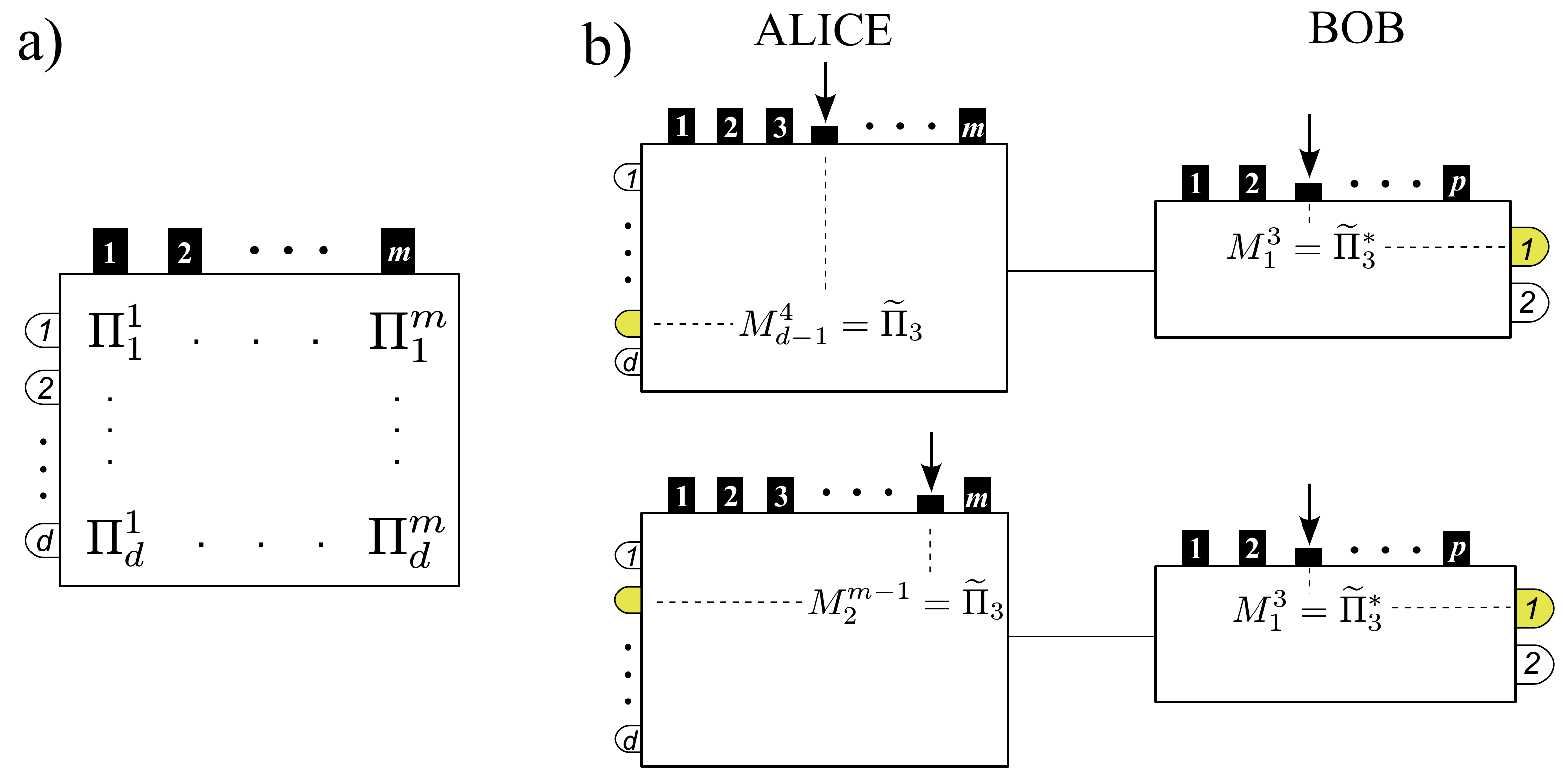}
 \caption{\textbf{Noncontextual assignments in the black-box
scenario.} (a) A KS proof consists of a single observer,
say Alice, who performs $m$ measurements of $d$ outcomes. The KS proof
requires that outcomes of different measurements correspond to the
same projector. There are altogether $p$ projectors, denoted by
$\tilde{\Pi}_j$, shared by different measurements. The common projectors impose constraints  that, if the outcomes are assigned by noncontextual deterministic maps, lead to contradictions.
(b) In the Bell test associated with the KS proof, Bob's box has $m_b=p$
possible measurements of $d_b=2$ outcomes. In the quantum setting, the
two observers share a maximally entangled state. Alice makes
$m_a=m$ measurements of $d_a=d$ outcomes, which correspond to the observables in the KS proof.
Bob's measurements are perfectly correlated with the $p$ projectors $\widetilde{\Pi}_{j}$
on Alice's side, thanks to the properties of the maximally entangled state. A local
model reproducing all these correlations would imply the
existence of a deterministic noncontextual model for Alice's
measurements, which is impossible.} \label{BlackBoxes}
\end{center}
\end{figure}

Let us now see how this highly nontrivial configuration of
measurements can be used to derive maximally nonlocal quantum
correlations. Consider the standard Bell scenario depicted in
Fig. \ref{BlackBoxes} (b). Two distant observers
(Alice and Bob) perform uncharacterized measurements in a
device-independent scenario. Let us assume that Alice can
choose among $m_a=m$ measurements of $d_a=d$ outcomes. On the other
hand, Bob can choose among $m_b=p$ measurements of $d_b=2$ outcomes,
labeled by $0$ and $1$. We denote Alice's (Bob's) measurement
choice by $x$ ($y$) and her (his) outcome by $a$ ($b$).
Collecting statistics at many instances of the experiment, they
compute the quantity $P(a,b|x,y)$, namely, the probability of
obtaining outcome $a$ and $b$ when measurements $x$ and $y$
were performed.

Consider next the following quantum realization of the experiment:
Alice and Bob perform their measurements on the bipartite
maximally entangled state $|\psi_d \rangle=\sum_{k=0}^{d-1}{1\over
\sqrt{d}}|kk\rangle$. When Alice chooses input $x$, measurement
$\{M^x_a=\Pi^z_i$, with $x=z$ and $a=i\}$ is performed. In turn,
when Bob chooses input $y$, the following measurement takes place:
$\{M^y_1=(\tilde\Pi_j)^*,M^y_2=\openone-(\tilde\Pi_j)^*$, with
$y=j\}$, where  the asterisk ($^*$) denotes complex conjugation. The properties
of $|\psi_d \rangle$  guarantee that these measurements by Alice
and Bob are perfectly correlated. Furthermore, they lead to the
nonsignalling value $\beta_{NS}$ of the following linear
combination of probabilities:
\begin{eqnarray}
 \label{bellgeneral}
 \nonumber\beta(P(a,b|x,y))&=&\sum_{y=1}^{p}\sum_{(a',x)\in D_{y}}[P(a=a',b=1|x,y)\\
&+& P(a\neq a',b=0|x,y)].
\end{eqnarray}
Indeed, for all the terms appearing in \eqref{bellgeneral},
$P(a=a',b=1|x,y)+P(a\neq a',b=0|x,y)=1$. This can be easily seen by
noticing that if Bob's output is equal to 1, Alice's system is
projected onto $\tilde{\Pi}_{y}=\Pi_{a'}^{x}$, and thus, the
result of Alice's measurement $x$ is $a'$. On the contrary, if Bob's box
outputs $0$, Alice's system is projected onto
$\openone-\tilde{\Pi}_{y}=\openone-\Pi_{a'}^x$, and thus,
Alice's outcome is such that $a\neq a'$. As the sum of the two
probabilities $P(a=a',b=1|x,y)$ and $P(a\neq a',b=0|x,y)$ can
never be larger than 1, one has
\begin{equation}
 \label{beta}
\beta_{Q}=\beta_{NS}\doteq\sum_{y=1}^{p}\sum_{(a',x)\in D_{y}}1.
\end{equation}

As for local correlations, we now show that it is
$\beta_{L}\leq\beta_{NS}-1$. To see this, recall first that the
maximum of \eqref{bellgeneral} over local models is always reached
by some deterministic model, in which a deterministic outcome is
assigned to every measurement [and all probabilities in
\eqref{bellgeneral} can thus only be equal to 0 or 1]. Hence,
deterministic models can only feature $\beta_L \in \mathbb{Z}$.
Therefore, it suffices to show that the maximum of
(\ref{bellgeneral}) over local models satisfies
$\beta_{L}<\beta_{NS}$. This can be proven by \emph{reductio ad
absurdum}. Suppose that a local deterministic model attains the
value $\beta_{NS}$. The model then specifies the outcomes $a$ and
$b$ on both sides for all measurements. Equivalently, it can be
understood as a definite assignment to every measurement outcome
on Alice's and Bob's sides: $v_{A}(M^x_a)\in \{1,0\}$ and
$v_{B}(M^y_b)\in \{1,0\}$, with $\sum_a v_A(M^x_a)=1=\sum_b
v_B(M^y_b)$, for all $x$ and $y$, respectively. If
(\ref{bellgeneral}) reaches its maximum algebraic value, the
assignment map is subject to the constraints
$v_{A}(M_{a}^{x})=v_{B}(M^y_1)=v_{A}(M_{a'}^{x'})$ for all $(a,x)$
and $(a',x')\in D_{y}$. Now, since $\{M_{a}^{x}\}$ is in
one-to-one correspondence with the projectors $\{\Pi_{i}^{z}\}$,
$v_{A}$ can then be thought of as a valid noncontextual
deterministic assignment map for $\{\Pi_{i}^{z}\}$. This, however,
is prohibited because $\{\Pi_{i}^{z}\}$ is a KS proof. Thus, one
concludes that $\beta_{L}\leq\beta_{NS}-1$.

The desired Bell inequality is then
\begin{equation}
 \label{bellineqgeneral}
\beta(P(a,b|x,y))\leq\beta_{NS}-1,
\end{equation}
with $\beta(P(a,b|x,y))$ defined by \eqref{bellgeneral} and
$\beta_{NS}$ defined by \eqref{beta}. This implies that the quantum
correlations obtained above from $|\psi_d\rangle$ feature $p_L=0$,
as they achieve the nonsignalling value of a Bell inequality,
which is in turn equal to its algebraic value.

Before concluding this section, we would like to emphasize that
this recipe can lead to other, possibly nonequivalent, Bell
inequalities. For instance, it is possible to keep Alice's
measurements equal to those in the KS proof and replicate them on
Bob's side, \ie, $\{M^y_b=(M^x_a)^*$, with $y=x$ and $b=a\}$. Note
that then all the projectors needed to enforce the KS constraints
on Alice's side by means of perfect correlations appear on Bob's
side. Other examples are provided by some proofs that possess
inherent symmetries, allowing for peculiar distributions of the
contexts in the proof between Alice's and Bob's sides, as is
discussed in the next section. 


\section{A SIMPLE BELL INEQUALITY}
\label{AsimpleBell}

The previous recipe is fully general. In this section, in contrast,
we apply the ideas just presented to derive a specific Bell
inequality maximally violated by quantum mechanics from one of the
most elegant KS proofs, introduced by Peres and Mermin
\cite{Mermin90,Peres91}. Apart from being one of the simplest Bell
inequalities having this property, its derivation shows how
symmetries in the KS proof can be exploited to simplify the
previous construction.


\begin{table}[t!]
\begin{center}
\begin{tabular}{|c|ccc|c|}
\hline\hline
& $y=1$ & $y=2$ & $y=3$ & \\
 \hline
$x=1$ & $Z_2$ & $X_1$ & $X_1 Z_2$ & $=\openone$ \\
$x=2$ & $Z_1$ & $X_2$ & $Z_1 X_2$ & $=\openone$ \\
$x=3$ & $Z_1 Z_2$ & $X_1 X_2$ & $Y_1 Y_2$ & $=-\openone$ \\
 \hline
 & $=\openone$ & $=\openone$ & $=\openone$ & $\prod$\\
\hline\hline
 \end{tabular}
 \end{center}
\caption{\label{Table1SI} {\bf The Peres-Mermin square}. One of
the simplest KS proofs was derived by Peres and Mermin
\cite{Peres91, Mermin90} and is based on the nine observables of
this table. The observables are grouped into six groups of three,
arranged along columns and rows. $X_n$, $Y_n$, and $Z_n$ refer to
Pauli matrices acting on qubits $n=1$ and $n=2$, which span a
four-dimensional Hilbert space. Each group constitutes a complete
set of mutually commuting (and therefore compatible) observables,
defining thus a context. In this way, there are six contexts, and
every observable belongs to two different ones. The product of all
three observables in each context is equal to the identity
$\openone$, except for those of the third row, whose product gives
$-\openone$. It is impossible to assign numerical values 1 or $-1$
to each one of these nine observables in a way that the values
obey the same multiplication rules as the observables. This, in
turn, implies that it is impossible to make a noncontextual
assignment to the 24 underlying projectors (not shown) in the
table (one common eigenbasis  per context, with four eigenvectors
each).}\end{table}

The Peres-Mermin (PM) KS proof is based on the set of observables
of Table~\ref{Table1SI}, also known as the PM square, which can
take two possible values, $\pm 1$. This proof in terms of
observables can be mapped into a proof in terms of 24 rank-1
projectors \cite{Peres91,CEG96}. To these projectors we could then
apply the formalism of the previous section  and derive Bell
inequalities maximally violated by quantum correlations of the
sort of \eqref{bellineqgeneral}. However, some special features of
this particular KS proof allow one to simplify the process and
derive a simpler inequality straight from the observables. The key
point is that in the PM square each operator appears in two
different contexts, one being a row and the other a
column. 
This allows one to distribute the contexts between Alice and Bob
in such a way that Alice (Bob) performs the measurements
corresponding to the rows (columns) (see also \cite{horo}). The
corresponding Bell scenario, then, is such that Alice and Bob can
choose among three different measurements $x,y\in \{1,2,3\}$ of
four different outcomes, $a,b\in \{1,2,3,4\}$. Consistent with
the PM square, we associate in what follows Alice and Bob's
observables $x$ and $y$ with the rows and columns
of the square, respectively, and divide the four-value outputs into two bits,
$a=(a_1,a_2)$ and $b=(b_1,b_2)$, each of which can take the
values $\pm 1$.

Consider first the following quantum realization: Alice and Bob
share two two-qubit maximally entangled states $|\psi_4\rangle ={
1 \over \sqrt{2}} (|00\rangle + |11\rangle )_{12}\otimes {1 \over
\sqrt{2} }(|00\rangle + |11\rangle )_{34}$, which is equivalent to
a maximally entangled state of two four-dimensional systems. Alice
possesses systems $1$ and $3$, and Bob possesses systems $2$ and $4$. Alice
can choose among three different measurements that correspond to
the three rows appearing in Table~\ref{Table1SI}. If Alice chooses
input $x$, the quantum measurement defined by observables placed
in row $x$ is performed. Note that the measurement acts on a four-dimensional quantum state; thus there exist four possible outcomes
(one for each eigenvector common to all three observables), which in our
scenario are decomposed into two dichotomic outputs. We define $a_i$
to be the value of the observable placed in column $y=i$ for
$i=1,2$. The value of the third observable in the same row is
redundant as it can be obtained as a function of the other two.
Equivalently, Bob can choose among three measurements that
correspond to the three columns appearing in Table~\ref{Table1SI}. If Bob
chooses input $y$, outputs $b_j$ are the values of observables
placed in column $y$ and row $x=j$ for $y=1,2,3$ and $j=1,2$. This
realization attains the algebraic maximum $\beta_Q=\beta_{NS}=9$
of the linear combination
\begin{eqnarray}
 \label{linearcomb}
\nonumber \beta&=&\langle a_{1} b_{1}|1,1\rangle + \langle a_{2} b_{1}|1,2\rangle + \langle a_{1} b_{2}|2,1\rangle \\
\nonumber &+&  \langle a_{2} b_{2}|2,2\rangle + \langle a_{1} a_{2} b_{1}|1,3\rangle
 + \langle a_{1} a_{2} b_{2}|2,3\rangle \\
\nonumber &+&\langle a_{1} b_{1} b_{2}|3,1\rangle + \langle a_{2} b_{1} b_{2}|3,2\rangle\\
&-& \langle a_{1} a_{2} b_{1} b_{2}|3,3\rangle,
\end{eqnarray}
where $\langle f(a_1,a_2,b_1,b_2)|x,y\rangle$ denotes the
expectation value of a function $f$ of the output bits
for the measurements $x$ and $y$.

To prove this statement, let us first focus on the term $\langle
a_1 b_1 |1,1 \rangle$. Bit $b_1$ is obtained as the
outcome of the measurement of the quantum observable $Z_4\otimes
\openone_2$. As the measurement is performed on the maximally
entangled state, the state on Alice's side is effectively
projected after Bob's measurement onto the eigenspace of
$Z_3\otimes \openone_1$ with eigenvalue $b_1$. Bit $a_1$ is
defined precisely as the outcome of the measurement of the
observable $Z_3\otimes \openone_1$; thus $a_1=b_1$ and $\langle
a_1 b_1 |1,1 \rangle=1$. The same argument applies to the
first four terms in~\eqref{linearcomb}. Consider now the term
$\langle a_1 a_2 \cdot b_1 |1,3 \rangle$. Bit $b_1$ is
the outcome of the measurement of the observable $Z_4\otimes X_2$.
The state after Bob's measurement is effectively projected on
Alice's side onto the eigenspace of $Z_3\otimes X_1$ with
eigenvalue $b_1$. Bit $a_1 \cdot a_2$ is obtained as the
 measurement output of the observable $Z_3\otimes X_1$; thus $a_1 \cdot
a_2=b_1$ and $\langle a_1 a_2 \cdot b_1 |1,3 \rangle=1$. The
same argument applies to the four terms involving products of
three bits. The last term $\langle a_1 a_2 \cdot b_1 \cdot
b_2 |3,3 \rangle$ requires a similar argument. Bit $a_1 \cdot
a_2$ is obtained as the output of the operator $Y_3\otimes Y_1$
(note that the product of the observables associated with $a_1$ and
$a_2$ is $Y_3\otimes Y_1$, see Table~\ref{Table1SI}). Thus the
state is effectively projected onto the eigenspace of $Y_4\otimes
Y_2$ with eigenvalue $a_1 \cdot a_2$. The bit $b_1 \cdot b_2$ is
precisely the meqasurement outcome of
 $-Y_4\otimes Y_2$, thus $a_1 \cdot a_2=-b_1
\cdot b_2$ and $\langle a_1 a_2 \cdot b_1 \cdot b_2 |3,3
\rangle=-1$.

We move next to the classical domain, to show that the maximum
value of polynomial \eqref{linearcomb} attainable by any local
model is $\beta_{L}=7$, and thus, the inequality
\begin{equation}
 \label{inequality}
\beta\leq7,
\end{equation}
with $\beta$ defined by \eqref{linearcomb}, constitutes a valid
Bell inequality, maximally violated by quantum mechanics.
Remarkably, this inequality has already appeared in Ref.
\cite{avncabello} in the context of all-versus-nothing
nonlocality tests. Computing the local bound $\beta_{L}=7$ can
easily be performed by brute force (that is, by explicitly
calculating the value of $\beta_{L}$ for all possible assignments).
However, it is also possible to derive it using  arguments similar
to those in the previous section. In the PM square, each of the
nine dichotomic observables belongs to two different contexts, one
being a row and the other a column, as mentioned. Therefore, nine
correlation terms are needed to enforce the KS constraints. As
said, the symmetries of the PM square allow one to split the
contexts between Alice and Bob, arranging these correlation terms
in a distributed manner. Such correlation terms correspond
precisely to the nine terms appearing in \eqref{linearcomb}.
Again, the existence of a local model saturating all these terms
would imply the existence of a noncontextual model for the PM
square, which is impossible.


\section{BOUND ON THE LOCAL CONTENT USING OTHER BELL INEQUALITIES}
\label{otherineqs}
The scope of this section is to show how the previous
construction offers important experimental advantages when
deriving bounds on the local content of quantum correlations.
First of all, and contrary to some of the examples of quantum
correlations with no local part~\cite{bkp}, the Bell inequalities
derived here not only involve a finite number of measurements but
are in addition resistant to noise. Moreover, as shown in what
follows, they allow one to obtain experimental bounds on the
nonlocal part that are significantly better than those based on
other Bell tests.

Let us first consider the Collins-Gisin-Linden-Massar-Popescu inequalities presented in \cite{CGLMP}. These inequalities
are defined for two measurements of $d$ outcomes. The maximal
nonsignalling violation of these inequalities is equal to
$\beta_{NS}=$4, while the local bound is $\beta_L=2$. The maximal
quantum violation of these inequalities is only known for small
values of $d$ \cite{ADGT,NPA}. A numerical guess for the maximal
quantum violation for any $d$ was provided in \cite{ZG}. This
guess reproduces the known values for small $d$ and tends to the
nonsignalling value when $d\rightarrow\infty$. Assuming
the validity of this guess, a bound on the local content
comparable to the experimental value reported in the next section,
namely, ${p_L}_{max}=0.218\pm0.014$, requires a number of outputs
of the order of 200 (see \cite{ZG}), even in the ideal noise-free
situation. Note that the known quantum realization attaining this
value involves systems of dimension equal to the number of
outputs, that is, 200, and the form of the quantum state is rather
complicated. If the quantum state is imposed to be maximally
entangled, the maximal quantum violation tends to $2.9681$, which
provides a bound on the local content of just
${p_L}_{max}\approx0.5195$.


\begin{figure}[t!]
\begin{center}
\includegraphics[width=0.95\columnwidth,keepaspectratio]{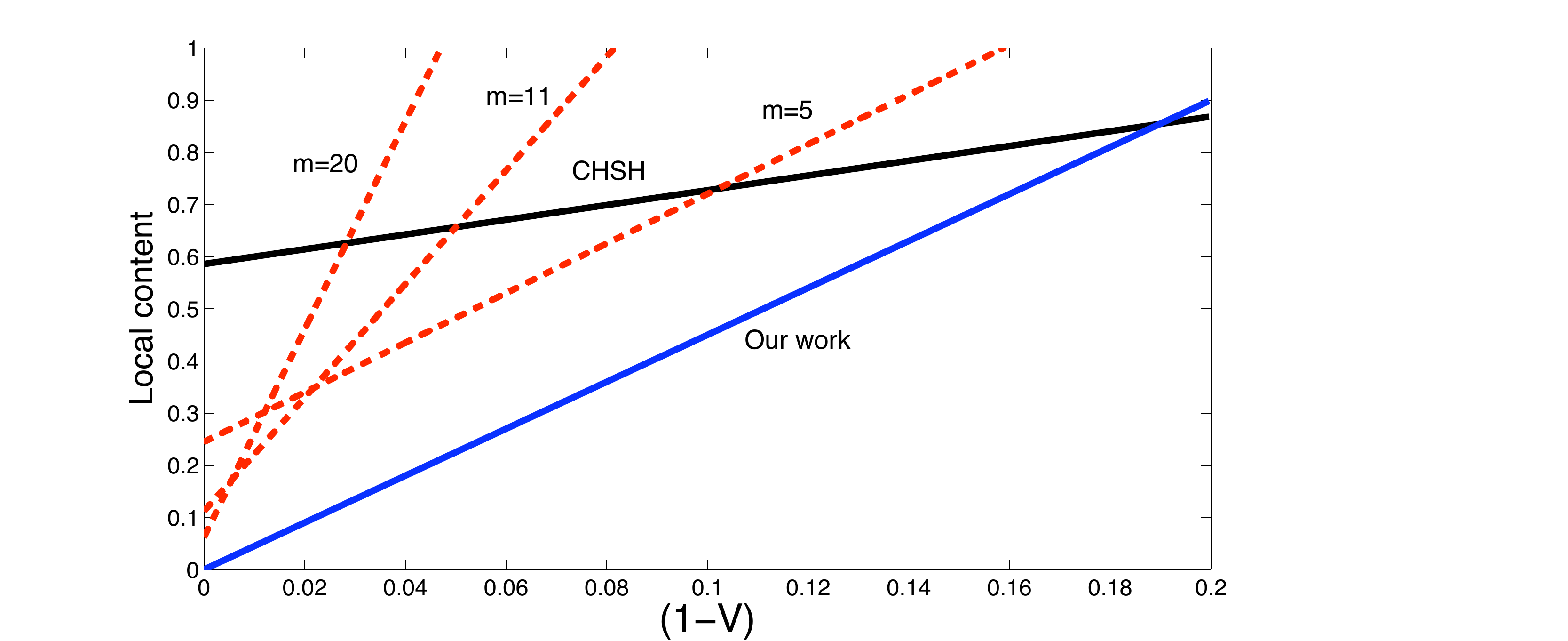}
\caption{\textbf{Resistance to noise of different Bell tests.} Dashed red
curves show the resistance to noise of the chained
inequality \cite{chained} for different numbers  $m$ of measurements.
The local content and also the resistance to noise tend to zero when
the number of measurements tends to infinity, as expected. Standard Bell inequalities,
such as the Clauser-Horne-Shimony-Holt (CHSH) inequality \cite{CHSH69},
can be violated in a robust manner and with few measurements, but the obtained bound on the
local content never goes to zero (in fact, the CHSH inequality is
the chained inequality \cite{chained} for $m=2$). Inequality \eqref{inequality}
(solid blue curve) in contrast combines all three features:
its violation is resistant to noise and requires few measurements, and its bound on the local content is equal
to zero in the noise-free case.} \label{Robustness}
\end{center}
\end{figure}

The chained inequalities \cite{chained, bkp}, defined in a scenario
where Alice and Bob can both perform $m$ measurements of $d$
outcomes, provide a bound on the local content that tends to
zero with the number of measurements,
$m\rightarrow\infty$ \cite{bkp}. However, in this limit the
nonlocality of the corresponding quantum correlations is not
resistant to noise (see Fig.~\ref{Robustness}), and thus, the
use of many measurements requires an almost-noise-free
realization. We compare the chained inequalities \cite{chained} for $d=2$ (the simplest case to implement) with our
inequality~\eqref{inequality} in a realistic noisy
situation. The quantum state is written as the mixture of the
maximally entangled state, as this state provides the maximal
quantum violation of both the chained inequality and
inequality~\eqref{inequality}, with white noise,
\begin{equation}
 \rho=V\proj{\psi_d}+(1-V)\frac{\openone}{d^2} .
\end{equation}
The amount of white noise on the state is quantified by $1-V$. The
bound on the local content then reads
${p_L}_{max}=\frac{\beta_{NS}-V\beta_{Q}+(1-V)\beta_{\openone}}{\beta_{NS}-\beta_{L}}$,
where $\beta_{\openone}$ is the value of the Bell inequality given
by white noise with the optimal measurements. We plot the obtained
results in Fig.~\ref{Robustness}. As shown there, the Bell
inequality considered here provides better bounds on the local
content than the chained inequalities for almost any value of the
noise.


\section{Experimental highly nonlocal quantum correlations}

\begin{figure}
 \begin{center}
  \includegraphics[width=8cm,angle=0]{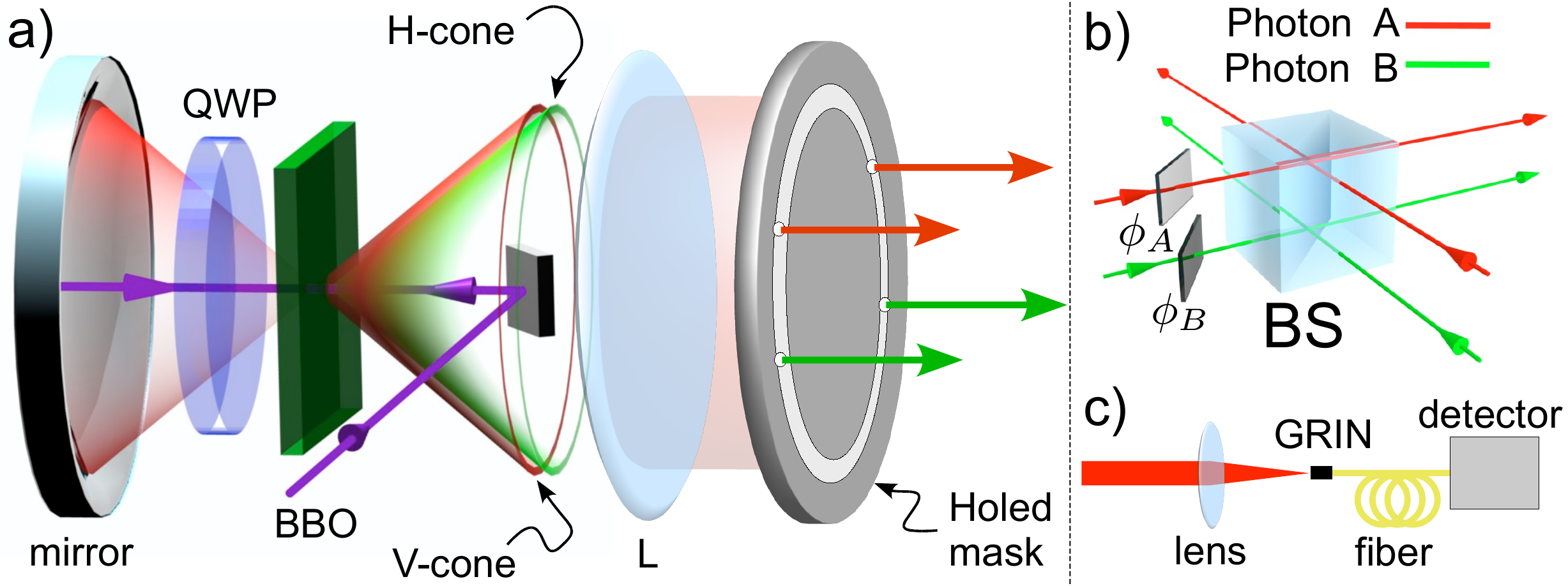}
\caption{\textbf{Experimental setup.} (a) Source of
hyperentangled photon states. The relative
phase between the states $\ket{HH}_{AB}$ and $\ket{VV}_{AB}$ can be varied by
translating the spherical mirror. A lens $L$ located at a focal
distance from the crystal transforms the conical emission into
a cylindrical one. (b) Scheme for the path measurements.
\textbf{c.} The parametric radiation is coupled into single-mode fibers by a GRIN
lens and sent to the detectors.}
\label{fig:setup}
\end{center}\end{figure}

We performed a test of  inequality \eqref{inequality} with two
entangled photons, $A$ and $B$, generated by spontaneous
parametric down conversion (SPDC). We used type-I phase matching
with a $\beta$-barium-borate (BBO) crystal. The source used a
single crystal and a double passage of the UV beam after the
reflection on a spherical mirror [see Fig. \ref{fig:setup} (a)] and generated the
hyperentangled state \cite{cine05prl}
\begin{eqnarray}
\nonumber \ket{\Psi}&=&\frac{1}{\sqrt2}({\ket{H}}_A{\ket{H}}_B+{\ket{V}}_A{\ket{V}}_B)\\
&&\otimes
 \frac{1}{\sqrt2}({\ket{r}}_A{\ket{l}}_B+{\ket{l}}_A{\ket{r}}_B)\,,
 \label{HE-pi-k}
\end{eqnarray}
where $\ket H$ ($\ket V$) represents the horizontal (vertical)
polarization and $\ket r$ and $\ket l$ are the two spatial path modes
in which each photon can be emitted. Maximally entangled state
$\ket{\psi_4}$ between $A$ and $B$, as defined in Sec.
\ref{AsimpleBell}, is recovered from \eqref{HE-pi-k} through the
following identification: $\ket{H}_{A,B}\equiv\ket{0}_{1,2}$,
$\ket{V}_{A,B}\equiv\ket{1}_{1,2}$,
$\ket{r}_{A}\equiv\ket{0}_{3}$, $\ket{l}_{A}\equiv\ket{1}_{3}$,
$\ket{l}_{B}\equiv\ket{0}_{4}$, and
$\ket{r}_{B}\equiv\ket{1}_{4}$. Therefore, state \eqref{HE-pi-k}
also allows for the maximal violation of \eqref{inequality}.

In the SPDC source, the BBO crystal is shined on by a vertically polarized continuous
wave (cw) Ar$^+$ laser ($\lambda_p$ = 364 nm), and the two
photons are emitted at degenerate wavelength $\lambda = 728$ nm and with horizontal polarization.
Polarization entanglement is generated by the double passage
(back and forth, after the reflection on a spherical mirror) of
the UV beam. The backward emission generates the so called
V cone: the SPDC horizontally polarized photons passing twice
through the quarter-wave plate (QWP) are transformed into
vertically polarized photons. The forward emission generates the
H cone [the QWP behaves almost as a half-wave plate (HWP) for the UV beam]. See Fig. \ref{fig:setup} (a). Thanks to temporal and spatial superposition, the
indistinguishability of the two perpendicularly polarized SPDC
cones creates polarization entanglement $({\ket{H}}_A{\ket{H}}_B+{\ket{V}}_A{\ket{V}}_B)/\sqrt{2}$.
The two
polarization entangled photons are emitted over symmetrical
directions belonging to the surface of the cone.
By selecting two pairs of correlated modes by a
four-holed mask \cite{cine05prl, vall08prl, chiu10prl} it is possible to generate
path entanglement.

\begin{figure*}[t!]
\begin{center}
\includegraphics[width=12cm,keepaspectratio]{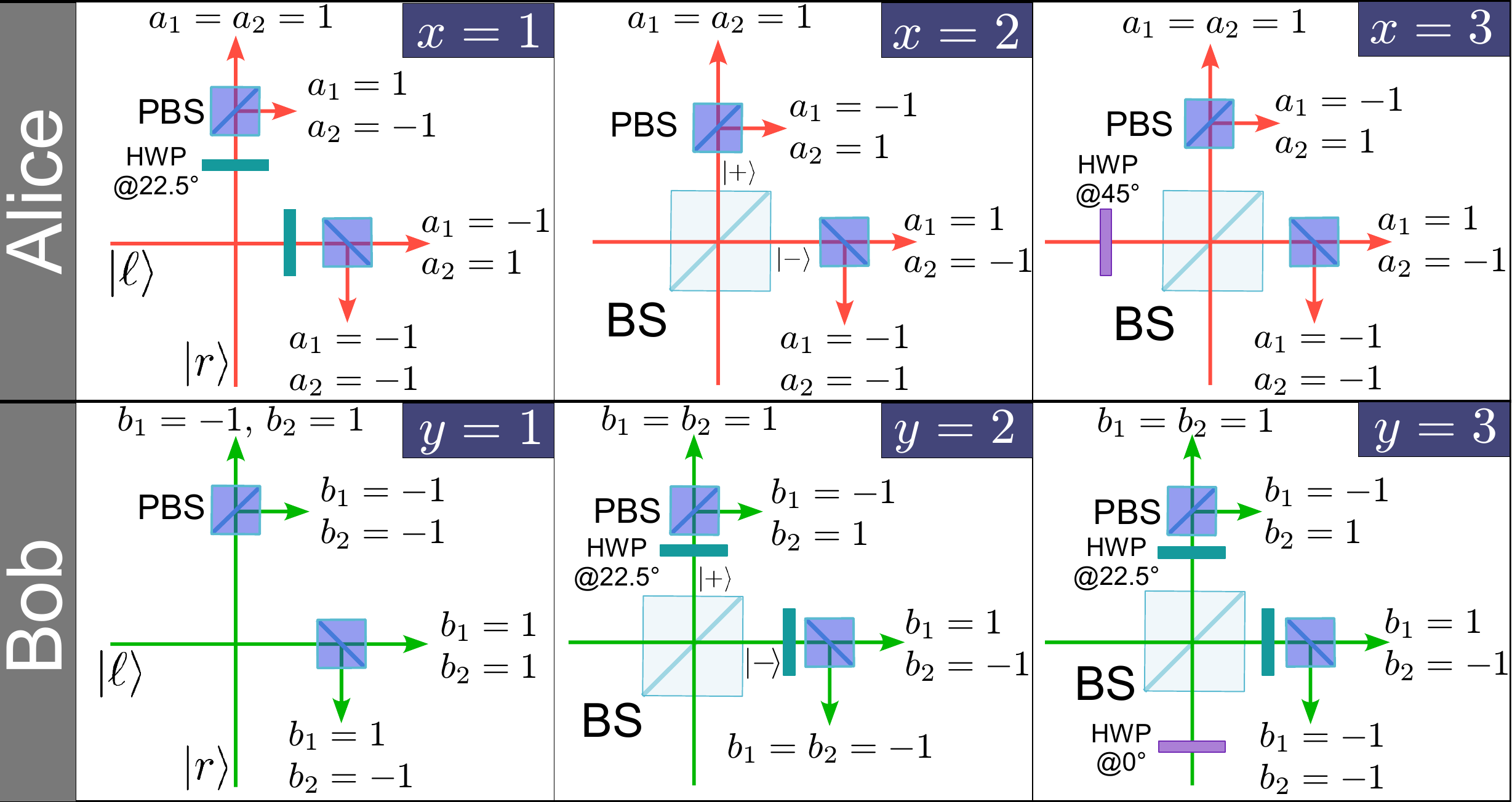}
\caption{\textbf{Measurement setups used by Alice and Bob.} See
text for a detailed explanation of the measurements. BS, beam
splitter; PBS, polarizing beam splitter; HWP, half-wave plate.}
\label{Analyzers}
\end{center}
\end{figure*}

In order to measure the path operators, the four modes of the
hyperentangled state are matched on a beam splitter (BS) in a
complete indistinguishability condition. This operation
corresponds to the projection onto
$\frac1{\sqrt2}(\ket{r}_{A}+e^{i\phi_{A}}\ket{l}_{A}) \otimes
\frac1{\sqrt2}(\ket{r}_{B}+e^{i\phi_{B}}\ket{l}_{B})$. Suitable
tilting of two thin glass plates allows one to set  phases
$\phi_{A}$ and $\phi_{B}$ [see Fig. \ref{fig:setup} (b)]. Photon
collection is performed by integrated systems of graded-index
lenses and single-mode fibers connected to single-photon counting
modules \cite{ross09prl, vall09apl}  [see Fig. \ref{fig:setup}
c)]. Polarization analysis is performed in each output mode by a
polarizing beam splitter (PBS) and a properly oriented HWP. The experimental setup used for each polarization
measurement setting is shown in Fig. \ref{Analyzers}.

\begin{table*}[t!]
\begin{center}
\begin{tabular}{c|c|c|c|c}
\multicolumn{5}{c}{Alice}\\
\hline\hline
&$a_1$=$-1$, $a_2$=$-1$&$a_1$=$-1$, $a_2$=$1$&$a_1$=$1$, $a_2$=$-1$&$a_1$=$1$, $a_2$=$1$\\
\hline
$x=1$&$\ket {-}\ket l$&$\ket {+}\ket l$&$\ket {-}\ket r$&$\ket {+}\ket r$\\
$x=2$&$\ket {V}\ket{-}$&$\ket {V}\ket{+}$&$\ket {H}\ket{-}$&
$\ket {H}\ket{+}$\\
$x=3$&$\ket{H}\ket{l}-\ket{V}\ket{r}$&$\ket{H}\ket{l}+\ket{V}\ket{r}$&$\ket{H}\ket{r}-\ket{V}\ket{l}$&$
\ket{H}\ket{r}+\ket{V}\ket{l}$\\
\hline\hline
\end{tabular}
\vskip1cm
\begin{tabular}{c|c|c|c|c}
\multicolumn{5}{c}{Bob}\\
\hline\hline
&$b_1$=$-1$, $b_2$=$-1$&$b_1$=$-1$, $b_2$=$1$&$b_1$=$1$, $b_2$=$-1$&$b_1$=$1$, $b_2$=$1$\\
\hline
$y=1$&$\ket {V}\ket r$&$\ket {H}\ket r$&$\ket {V}\ket l$&$\ket {H}\ket l$\\
$y=2$&$\ket {-}\ket{-}$&$\ket {-}\ket{+}$&$\ket {+}\ket{-}$&
$\ket {+}\ket{+}$\\
$y=3$&$\ket{+}\ket{r}-\ket{-}\ket{l}$&$\ket{+}\ket{r}+\ket{-}\ket{l}$&$\ket{-}\ket{r}-\ket{+}\ket{l}$&$\ket{-}\ket{r}+\ket{+}\ket{l}$\\
\hline\hline
\end{tabular}
\end{center}
\caption{ \label{tableII}
{\bf Measurement settings}. Each row
represents a measurement (context).
The four states in each row represent the four projectors of each measurement.
$a_{1,2}$ and $b_{1,2}$ are the two-bit outcomes of Alice and Bob respectively.
In each state, the first ket refers to polarization, while
the second one refers to path. $\ket\pm$ correspond to $\frac{1}{\sqrt2}(\ket H\pm\ket V)$ or
$\frac{1}{\sqrt2}(\ket r\pm\ket l)$, for polarization or path respectively.}
\end{table*}

The nine terms of Bell polynomial \eqref{linearcomb} correspond to
the different combinations between one of Alice's three contexts
and one of Bob's three contexts listed in Table \ref{tableII}. In
the settings $x=1,2$ ($y=1,2$) Alice (Bob) must project into
states that are separable between path and polarization
(eigenstates of Pauli operators $X$ and $Z$). To project into
$\{\ket{r},\ket{l}\}$ the modes are detected without BS. On the
other hand, the BS is used to  project into
$\frac1{\sqrt2}(\ket{r}\pm\ket{l})$. PBSs and wave plates have been
exploited to project into $\{\ket{H},\ket{V}\}$ or
$\frac1{\sqrt2}(\ket{H}\pm\ket{V})$. More details are needed for
contexts $x,y=3$, corresponding to the projection into single-photon Bell states (the two entangled qubits of the Bell state are
encoded in polarization and path of the single particle, see
Table \ref{tableII}). For instance, let us consider the projection
on the states $\ket{H}\ket{l}\pm\ket{V}\ket{r}$ and
$\ket{V}\ket{l}\pm\ket{H}\ket{r}$ for Alice. By inserting a HWP
oriented at $45^{\circ}$ on the mode $\ket{l}_{A}$ before the BS,
the previous states become $\ket{V}\ket{\pm}$ and
$\ket{H}\ket{\pm}$, respectively. The two BS outputs allow one to
discriminate between $\ket{r}+\ket{l}$ and $\ket{r}-\ket{l}$, while
the two outputs of the PBSs discriminate $\ket H$ and $\ket V$.

\begin{table}[t!]
\begin{center}
\begin{tabular}{cc}
\hline\hline
 Correlation & Experimental result\\
\hline $\langle a_1 b_1|1,1\rangle$ & $0.9968\pm0.0032$
\\
$\langle a_1 b_2|2,1\rangle$ & $0.9759\pm0.0058$
\\
$\langle a_2 b_1|1,2\rangle$ & $0.9645\pm0.0068$
\\
$\langle a_2 b_2|2,2\rangle$ & $0.941\pm0.010$
\\
$\langle a_1 a_2 b_1|1,3\rangle$ & $0.9705\pm0.0048$
\\
$\langle a_1 a_2 b_2|2,3\rangle$ & $0.9702\pm0.0049$
\\
$\langle a_1 b_1 b_2|3,1\rangle$ & $0.9688\pm0.0073$
\\
$\langle a_2 b_1 b_2|3,2\rangle$ & $0.890\pm0.013$
\\
$\langle a_1 a_2 b_1 b_2|3,3\rangle$ &
$-0.888\pm0.018$
\\
\hline\hline
\end{tabular}
\end{center}
\caption{\label{Table2b} {\bf Experimental results}. Errors were
calculated by propagating Poissonian errors of the counts.}
\end{table}

Table~\ref{Table2b} provides the experimental values of all nine
correlations in Bell polynomial \eqref{linearcomb}. The obtained
violation for Bell inequality \eqref{inequality} is
$\beta^{exp}_Q=8.564\pm0.028$ and provides the upper bound
${p_L}_{max}=0.218\pm0.014$. At this point it is important to
mention that another experimental test of \eqref{inequality} was
reported in Ref. \cite{pan} in the framework of all-versus-nothing
nonlocality tests. The violation in Ref. \cite{pan} is compatible
(within experimental errors) with the value obtained by our
experiment.

\section{Conclusions and discussions}


In this work we have provided a systematic recipe for
obtaining bipartite Bell inequalities from every proof of the
Kochen-Specker theorem. These inequalities are violated by
quantum correlations in an extremal way, thus revealing the fully
nonlocal nature of quantum mechanics. We have  shown that
these inequalities allow establishing experimental bounds on the
local content of quantum correlations that are significantly
better than those obtained using other constructions. This enabled
us to experimentally demonstrate a Bell violation leading to the
highly nonlocal bound $p_L\lesssim\ 0.22$.


The local content $p_L$ of some correlations $P(a,b|x,y)$ can be
understood as a measure of their locality, as it measures the
fraction of experimental runs admitting a local-hidden-variable
description. As mentioned, some of the previously known
examples of bipartite inequalities featuring fully nonlocal
correlations, i.e., $p_L=0$, for arbitrary dimensions require an
infinite number of measurement settings and are not robust against
noise \cite{epr2,bkp}. More standard Bell inequalities using
a finite number of measurements, such as the well-known
Clauser-Horne-Shimony-Holt inequality~\cite{CHSH69}, give a
local weight significantly larger than zero even in the noise-free
situation. Thus, the corresponding experimental violations,
inevitably noisy, have only managed to provide bounds on the local
content not smaller than 0.5 (see Table~\ref{Table3}). In
contrast, the theoretical techniques provided in this work enable
the experimental demonstration of highly nonlocal correlations.
This explains why the experimental bound provided in this work is
significantly better than those of previous Bell tests, even
including multipartite ones. In fact, multipartite
Greenberger-Horne-Zeilinger  tests \cite{ghz}  also in principle
yield $p_L=0$ \cite{bkp} using a finite number of measurements and
featuring robustness against noise. Still, to our knowledge, the
reported  experimental violations lead to significantly worse
bounds on $p_L$ (see Table~\ref{Table3}). Our analysis, then,
certifies that, in terms of local content, the present bounds allow a higher degree of nonlocal correlations than those reported in \cite{ADR82,WJSWZ98,KSWTGUW05,ZYCZZP03,PBSRG11} or in any other previous experiment of our knowledge.

\begin{table}[t!]
\begin{center}
\begin{tabular}{cc}
\hline\hline
 Experiment & $p_L$ \\
\hline
 Aspect \emph{et al.} \cite{ADR82} & $\lesssim\ 0.80$ \\
 Weihs \emph{et al.} \cite{WJSWZ98} & $\lesssim\ 0.64$ \\
 Kiesel \emph{et al.} \cite{KSWTGUW05} & $\lesssim\ 0.64$ \\
 Zhao \emph{et al.} \cite{ZYCZZP03} & $\lesssim\ 0.60$ \\
 Pomarico \emph{et al.} \cite{PBSRG11} & $\lesssim\ 0.49$ \\
 This work (and Yang \emph{et al.} \cite{pan}) & $\lesssim\ 0.22$ \\
\hline\hline
\end{tabular}
\end{center}
\caption{\label{Table3} {\bf Bounds on the local content of
quantum correlations from previous Bell experiments}. The
selection includes representative experiments testing different
forms of nonlocality, or Bell inequalities, in both  the
bipartite \cite{CHSH69,chained} and multipartite
 \cite{Ardehali92,SASA05} scenarios. Other published experiments,
not shown in the table, lead to
${p_L}_{max}> 0.49$. Note the significant improvement  given by the techniques discussed in this
work (see also Sec. \ref{otherineqs}).}
\end{table}

\begin{acknowledgments}
We acknowledge support from  Spanish projects 
FIS2008-05596 and FIS2010-14830,  QOIT (Consolider Ingenio
2010), a Juan de la Cierva grant, the European EU FP7 Project
Q-Essence, EU-Project CHIST-ERA QUASAR, EU-Project CHIST-ERA DIQIP, an ERC Starting Grant PERCENT, CatalunyaCaixa,
Generalitat de Catalunya,  Italian projects PRIN 2009 of Ministero dell'Istruzione, dell'Universitö e della Ricerca and FARI 2010 Sapienza
Universit\`{a} di Roma, and the Wenner-Gren Foundation.
\end{acknowledgments}

\end{document}